\documentclass[lettersize,journal]{IEEEtran}
\usepackage{amsmath,amssymb,amsfonts}
\usepackage{algorithmic}
\usepackage{graphicx}
\usepackage{subcaption}
\usepackage{textcomp}
\usepackage[table]{xcolor}
\usepackage{multirow}
\usepackage[misc]{ifsym}
\usepackage{url} 
\usepackage{booktabs}
\usepackage[numbers,sort&compress]{natbib}
\usepackage{tikz}
\usepackage{circledsteps}
\usepackage{listings}
\usepackage{soul}
\usepackage{xspace}
\usepackage{soul}
\usepackage[skins,breakable]{tcolorbox}
\usepackage{url}
\usepackage[autostyle]{csquotes}  
\usepackage{todonotes}
\usepackage{pgfplots}
\usepackage{pgfplotstable}
\usepackage{booktabs}
\usepackage{threeparttable}
\usepackage{tabularx}
\usepackage{hyperref}

\newcommand*\circled[2]{%
  \definecolor{currentcolor}{HTML}{#1}
  \tikz[baseline=(C.base)]
    \node[
      draw, 
      circle, 
      fill=currentcolor, 
      text=white, 
      inner sep=1pt
    ](C) {#2};\!
}

\newcommand{\fix}[1]{#1}

\newcommand{\minor}[1]{#1}

\setuptodonotes{inline}

\definecolor{coolblack}{rgb}{0.0, 0.18, 0.39}

\definecolor{formalshade}{rgb}{1.0,1.0,1.0}
\definecolor{side}{rgb}{0.0,0.2,0.6}

\definecolor{grayrow}{rgb}{0.7,0.7,0.7}
\definecolor{my4grey}{HTML}{525252}
\definecolor{my3grey}{HTML}{969696}
\definecolor{my2grey}{HTML}{cccccc}
\definecolor{my1grey}{HTML}{f7f7f7}

\newcommand{\RqOne}{\textbf{RQ1. To what extent does cross-project flakiness exist?}\xspace}

\newcommand{\RqTwo}{\textbf{RQ2. How do test scopes influence the likelihood of cross-project flakiness?}
\xspace}

\newcommand{\RqThree}{\textbf{RQ3.  What are the causes of inconsistent flakiness?}\xspace}

\usepackage{amsthm}

\usepackage[strict]{changepage}
  \definecolor{ABlue}{HTML}{127bca}
 \definecolor{LHScolor}{HTML}{555555}
\usepackage{framed}

\definecolor{formalshade}{rgb}{1.0,1.0,1.0}
\definecolor{side}{rgb}{0.0,0.2,0.6}

\newenvironment{formal}{%
  \MakeFramed{\advance\hsize-\width\FrameRestore}%
  \noindent\hspace{-4.55pt}
  \begin{adjustwidth}{}{7pt}%
  \vspace{2pt}\vspace{2pt}%
}
{%
  \vspace{2pt}\end{adjustwidth}\endMakeFramed%
}

\def\BibTeX{{\rm B\kern-.05em{\sc i\kern-.025em b}\kern-.08em
    T\kern-.1667em\lower.7ex\hbox{E}\kern-.125emX}}

\begin{document}

\title{Cross-Project Flakiness: A Case Study of the OpenStack Ecosystem}

\author{Tao Xiao, Dong Wang~\Letter, Shane McIntosh, Hideaki Hata, Yasutaka Kamei
\IEEEcompsocitemizethanks{\IEEEcompsocthanksitem{T. Xiao, Y. Kamei are with Kyushu University, Japan.

E-mail: \{xiao, kamei\}@ait.kyushu-u.ac.jp}
\IEEEcompsocthanksitem{D. Wang is with Tianjin University, China.

E-mail: dong\_w@tju.edu.cn}
\IEEEcompsocthanksitem{S. McIntosh is with the University of Waterloo, Canada.

E-mail: shane.mcintosh@uwaterloo.ca}
\IEEEcompsocthanksitem{H. Hata is with Shinshu University, Japan.

E-mail: hata@shinshu-u.ac.jp}}
}

\markboth{Journal of \LaTeX\ Class Files,~Vol.~18, No.~9, September~2020}%
{How to Use the IEEEtran \LaTeX \ Templates}

\maketitle

\begin{abstract}
Automated regression testing is a cornerstone of modern software development, often contributing directly to code review and Continuous Integration (CI). Yet some tests suffer from flakiness, where their outcomes vary non-deterministically. Flakiness erodes developer trust in test results, wastes computational resources, and undermines CI reliability. While prior research has examined test flakiness within individual projects, its broader ecosystem-wide impact remains largely unexplored. In this paper, we present an empirical study of test flakiness in the OpenStack ecosystem, which focuses on (1) cross-project flakiness, where flaky tests impact multiple projects, and (2) inconsistent flakiness, where a test exhibits flakiness in some projects but remains stable in others. By analyzing 649 OpenStack projects, we identify 1,535 cross-project flaky tests and 1,105 inconsistently flaky tests. We find that cross-project flakiness affects 55\% of OpenStack projects and significantly increases both review time and computational costs. Surprisingly, 70\% of unit tests exhibit cross-project flakiness, challenging the assumption that unit tests are inherently insulated from issues that span modules like integration and system-level tests. Through qualitative analysis, we observe that race conditions in CI, inconsistent build configurations, and dependency mismatches are the primary causes of inconsistent flakiness. These findings underline the need for better coordination across complex ecosystems, standardized CI configurations, and improved test isolation strategies.
\end{abstract}

\section{Introduction}

\IEEEPARstart{D}{evelopers} rely on automated regression testing to assess whether key functionality is perturbed by their changes. Ideally, each test failure should serve as a clear indication of problems in a change set; however, failures may occur due to non-deterministic factors. This behavior, known as \textbf{test flakiness}~\cite{luo2014empirical}, provides a misleading signal to developers regarding their changes. Flakiness arises when a test produces both passing and failing results when re-executed under identical conditions. Prior research has highlighted issues caused by test flakiness, such as wasted developer effort in investigating false positives~\cite{7332456, 10.1145/2610384.2610404}, reduced efficiency of Continuous Integration (CI) systems~\cite{durieux2020empirical,lam2020study}, and the obfuscation of software defects~\cite{10.1145/2610384.2610404, vahabzadeh2015empirical}.

Addressing test flakiness is a persistent challenge in software testing.
For example, Google has documented that between 2\% and 16\% of its computational resources are allocated to re-execute flaky tests \cite{micco2017state}. 
A survey of 58 Microsoft developers found that they ranked test flakiness as the second most important perceived barrier to software deployment \cite{lam2019root}. 
The economic impact of test flakiness is also substantial. Herzig et al. \cite{herzig2015art} estimated that the cost of investigating test failures caused by flakiness at Microsoft amounts to approximately \$7.2 million for a single product. 
Additionally, an analysis of GitHub commits revealed that ``1 in 11 commits had at least one [failure] caused by a flaky test'' \cite{ghblog}. 
Similar issues have been reported by Mozilla~\cite{mozilla} and Huawei~\cite{jiang2017causes}, further indicating the widespread impact of test flakiness. 

Prior research has predominantly focused on flakiness at the programming level~\cite{parry2021survey}. To the best of our knowledge, the cascading impact of flakiness at the ecosystem level, which encompasses a set of socio-technically interdependent software projects~\cite{manikas2013software,mens2017towards,jansen2013defining}, has not yet been explored. While individual projects within the ecosystem are managed by their respective project teams or organizations, the strength of the ecosystem stems from the social interactions among these teams (both within and across projects~\cite{mens2011analysing,de2016social}) as well as their ability to foster reuse functionalities through technical dependencies~\cite{adams2016empirical,bogart2016break,decan2017empirical,kikas2017structure}. Flakiness in such an ecosystem presents unique challenges since it does not remain insulated within a single project. Instead, it propagates through shared dependencies and interfaces that multiple projects rely on.

In this paper, we perform a case study of OpenStack, a renowned infrastructure-based ecosystem~\cite{foundjem2021release} developed by 50 participating industrial members, including large-scale and well-known companies, such as AMD, Microsoft, and Cisco. Through our preliminary exploration, we identify \textbf{cross-project flakiness} and \textbf{inconsistent flakiness} phenomena in the OpenStack ecosystem (detailed in Section~\ref{motivate}). Cross-project flakiness occurs when a flaky test affects multiple projects. Inconsistent flakiness refers to instances where a test exhibits non-deterministic behavior in a subset of projects while remaining stable in others.
While prior research~\cite{luo2014empirical, maipradit2023repeated} has explored test flakiness within individual projects, its characteristics and impact on multiple projects, particularly at the ecosystem level, remain largely unexplored. More specifically, we address the following three Research Questions (RQs):

\smallskip
\noindent
\RqOne 
\indent \textbf{Motivation.} The prevalence of cross-project flakiness in the OpenStack ecosystem remains unclear, highlighting a critical knowledge gap given the limited resources available to address the issue.
Therefore, we first set out to quantify the prevalence of cross-project flakiness.
Providing a foundational understanding of its behavior for assisting team leads of software ecosystems, responsible for overseeing all technical aspects and representing contributors, in making informed decisions for effective intervention.\\
\indent \textbf{Results.} We find that cross-project flakiness impacts a large proportion (55\%) of OpenStack projects, occurring with greater frequency than flakiness confined to individual projects. Moreover, cross-project flakiness has rapidly accumulated over time in OpenStack. 
These findings are a call to action for addressing cross-project flakiness to enhance CI reliability
and reduce resource consumption.


\smallskip
\noindent
\RqTwo \\
\indent \textbf{Motivations.} Software testing can be categorized into various types to comprehensively validate the software, such as unit tests, functional tests, and integration tests.
Each test type may interact with system dependencies and infrastructure in unique ways, which could potentially lead to different degrees of flakiness (e.g., number of projects being affected by the flaky test) in interconnected ecosystems like OpenStack.
By exploring how these various test types contribute to cross-project flakiness, researchers and developers can pinpoint which test types are most vulnerable to inducing such instability. Consequently, team leads of software ecosystems could leverage these insights to refine existing test design strategies and proactively address identified flakiness phenomena. \\
\indent \textbf{Results.} We find that the two flakiness phenomena are prevalent in the OpenStack ecosystem, with 1,535 and 1,105 flaky tests presenting with cross-project flakiness and inconsistent flakiness, respectively. Unit tests are the most likely to exhibit flakiness, accounting for 7,202 flaky tests observed across 30,550 test executions. Moreover, we observe a strong correlation between test scopes (i.e., unit, integration, system, scenario, or API tests) and flakiness levels (i.e., the extent to which flakiness propagates across the network of projects). We find that 64\% of API tests and 41\% of scenario tests are prone to causing flakiness across multiple OpenStack projects. Surprisingly, cross-project flakiness affects 70\% of unit tests, suggesting that even tests that are traditionally well-isolated can contribute to ecosystem-wide flakiness. These findings suggest that test strategies may benefit from refinement, especially in terms of test isolation and resource management, to mitigate flakiness at its root cause.


\smallskip
\noindent
\RqThree  
\indent \textbf{Motivations.}  In this RQ, we conduct an in-depth analysis to uncover the underlying causes of inconsistent flakiness, that is, why certain projects experience flakiness under similar conditions while others do not. Understanding these causes not only helps in developing targeted mitigation strategies but also guides the refinement of testing practices and tools for interconnected ecosystems, ultimately leading to more stable and robust software systems benefiting all stakeholders. \\
\indent \textbf{Results.} We find that race conditions in the CI process are the predominant cause of inconsistent flakiness, accounting for \fix{89}\% of cases. Additionally, \fix{both} mismatched CI configurations and dependency management issues contribute to 21\% of cases. These observations call for standardized CI configurations and enhanced coordination among OpenStack projects to reduce inconsistencies.

\medskip
To understand how OpenStack developers perceive our empirical observations, we solicit their feedback through a questionnaire. We received valid responses from 15 contributors, core developers, and project maintainers. The respondents explain that OpenStack stakeholders have also encountered the two flakiness phenomena. Moreover, the respondents commented on and proposed mitigation strategies, such as the shift from the ``recheck and wait'' paradigm to early intervention on flaky tests or standardizing CI configuration and environment. 

To facilitate replication and promote future work, we have made a replication package publicly available~\cite{xiao25flaky}, which includes our raw data set, manually labelled data, anonymized questionnaire responses, and analysis code scripts.

\begin{figure*}[t]
    \centering
    \includegraphics[width=\textwidth]{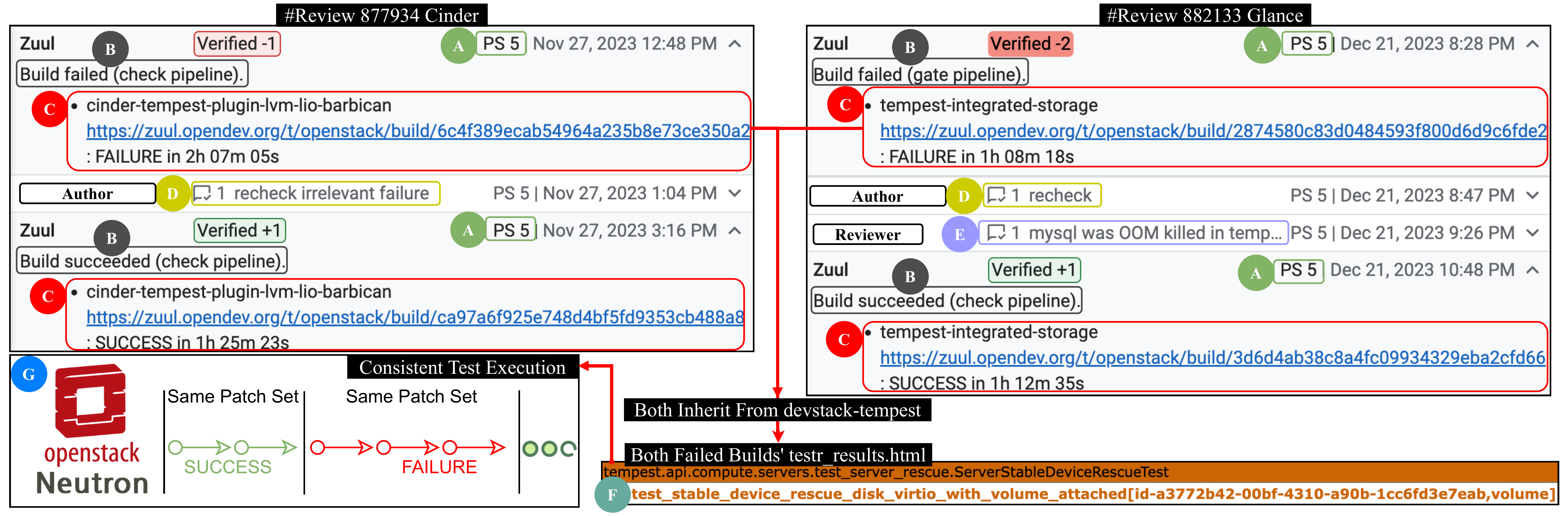}
    \caption{Real-world example of cross-project flakiness during the code review \#877934 (Cinder project) and \#882133 (Glance project) from OpenStack community.}
    \label{fig:exm}
\end{figure*}

\section{Code Review and Continuous Integration}
To investigate test flakiness at the ecosystem level, we explain its occurrence in practice, particularly in OpenStack.
Modern code review is a lightweight process widely adopted in both industrial and open-source settings~\cite{wang2021can, rigby2013convergent}, conducted online and asynchronously through review tools such as Gerrit.\footnote{\url{https://www.gerritcodereview.com/}}
The process typically commences when a code author submits a set of changes to a code base, along with a description of the changes, to the review tool. Then, reviewers critique the premise, content, and structure of the submitted changes, providing feedback for the author to refine their work.

While authors may test their changes locally before submission, not all developers have the necessary resources to run the complete test suite. Consequently, reviewers cannot assume that all relevant tests have been conducted. 
To address this limitation, the code review process is tightly integrated with CI through bots (e.g., the Zuul bot of OpenStack) that automatically perform builds and run regression tests. Along with reviewers' feedback, CI bots report the results of builds and tests.
Authors may upload revised change sets in response to reviewer feedback and to satisfy CI bots. 
This cycle of review, testing, and revision continues until the latest revision meets both the CI bots' criteria and the reviewers' expectations. 
Throughout this paper, in accordance with \path{Gerrit} terminology, a \textit{change set} is referred to as a \textit{code review}, and each \textit{revision of a change set} is referred to as a \textit{patch set}.




\section{Real-World Example of Cross-Project Flakiness} 
\label{motivate}

Figure~\ref{fig:exm} presents a real-world example of cross-project flakiness observed during code review \path{#877934}\footnote{\url{https://review.opendev.org/c/openstack/cinder/+/877934}} in Cinder and \path{#882133}\footnote{\url{https://review.opendev.org/c/openstack/glance/+/882133}} in Glance from the OpenStack ecosystem. 
In this example, 
\circled{82B366}{A} denotes the patch set on which the Zuul build was based, while \fix{\circled{4D4D4D}{B} represents the build status and the triggering pipelines.
\circled{FF0000}{C} refers to the build jobs executed by Zuul, while \circled{CCCC00}{D} represents the \path{recheck} (a re-run technique to mitigate non-deterministic behavior) request invoked by the author. 
\circled{9999FF}{E} indicates the review comment explaining the random failure provided by the patch reviewer, while both failing builds pinpoint to the same failing test (\circled{67AB9F}{F}). Finally, this test can also be executed in other projects, shown as \circled{007FFF}{G}.}

\fix{OpenStack's Zuul CI workflow is event-driven, where new changes trigger different pipelines based on their review status, such as \textit{check} pipeline (when uploading a new patch set) or \textit{gate} pipeline (before merging changes). Once the \textit{gate} has failed, the submitted patch set must proceed through a \textit{check} pipeline similar to \circled{4D4D4D}{B} (transitioning from \textit{gate} to \textit{check}) on the right side of Figure~\ref{fig:exm}. In OpenStack, build jobs and test suites could be shared across different projects through inheritance from common job definitions. For example, both \circled{FF0000}{C} jobs from two reviews inherit from common parents, \path{devstack-tempest} in the Tempest project. These two jobs in Cinder and Glance executed the same set of standard Tempest API tests to ensure that a modification in one project does not break the OpenStack API.}

\fix{As shown in the figure, the authors of both changes }request a rerun of the CI job after its failure by initiating a \path{recheck} command (\circled{CCCC00}{D}).
The two different statuses of the two CI builds, ``SUCCESS'' and ``FAILURE,'' are shown in \circled{FF0000}{C}. These statuses are from the Zuul CI, and both jobs are based on patch set 5 (\circled{82B366}{A}). \fix{In \circled{FF0000}{C}}, we refer to the failing CI builds as the \textbf{flaky build}, the unique identifier \fix{in the URL} as the identifier; and the failing CI jobs as \textbf{flaky job}, the name of the CI jobs as the identifier. In this figure, both flaky builds wasted CI time or exhausted CI resources for more than one hour. The bottom portion of this figure illustrates that the failing test (\circled{67AB9F}{F}) caused the failure of the flaky build, which we refer to as the \textbf{flaky test}. Although the flaky builds originated from two different reviews and projects, both pointed to the same flaky test. We define \fix{it} as \textbf{cross-project flakiness}\fix{, which refers to \textit{non-deterministic test failures that manifest across multiple projects in a software ecosystem}. It serves as a signal to the tests for their instability or non-deterministic behavior in this ecosystem.} This particular flaky \fix{API} test\fix{~\cite{apitest1}, asserts that the attached volume is still present and accessible in rescue mode,} is the most frequently occurring in our data set, accounting for 201 instances. It has not only been flaky in the Cinder and Glance projects, but also in 14 other OpenStack projects, such as Nova, Devstack, and Tempest. \fix{According to the bug report,\footnote{\url{https://bugs.launchpad.net/nova/+bug/1960346}} Nova fails to detach the volume via Libvirt during test cleanup, despite retries. This flakiness is suspected to be attributed to changes introduced in Libvirt.} 

We investigate cross-project flakiness along flaky builds (coarse-grained level) and flaky tests (fine-grained level). \fix{In this example, the flaky test (\circled{67AB9F}{F}) can be consistently executed in the Neutron project (\circled{007FFF}{G}). This test exhibits that \textit{a deterministic status for other projects, while flakiness for one or more projects in a software ecosystem},
we define such inconsistent behavior as \textbf{inconsistent flakiness}. It serves as a signal to tests that the non-deterministic behavior may not be only inherited by the test logic (true behavioral inconsistency), but could also be triggered by configurations or environments in the project.}




\section{Case Study Design}
\label{sec:cs}
In this section, we describe the selection of our studied ecosystem and our approach to data preparation.

\subsection{Studied Ecosystem}
\label{sec:se}
Our study aims to investigate the flakiness and its propagation across multiple interconnected projects at the ecosystem level. Thus, we need to collect testing logs from projects with frequently occurring builds on identical code (flakiness) in CI. Below, we illustrate our rationale for selecting OpenStack.

\textbf{Popularity of the Studied Ecosystem.} 
OpenStack is one of the largest open-source communities, with approximately 13 million lines of code contributed by around 12,000 developers. 
This ecosystem has been extensively studied within the code review context~\cite{mcintosh2016empirical, wang2021understanding, maipradit2023repeated}.
OpenStack comprises open-source projects supported by established organizations and companies (e.g., AMD, Microsoft, and Cisco), collaboratively developing a cloud computing platform.\footnote{\url{https://www.openstack.org/community/supporting-organizations/}} Over the past decade, the OpenStack community has dedicated substantial effort to documenting and implementing code reviews, making it a valuable source of insights for similar communities. 

\textbf{Significance of the Studied Ecosystem.}
Red Hat~\cite{redhat2015} advocates for the ``Upstream First'' policy, while the OpenStack community~\cite{seeding2018} supports ``Long-Term Support'' and ``Extended Maintenance'' to maintain historically stable branches instead of terminating them, thereby providing a cost-effective and sustainable model for innovation in open-source platforms.
These policies may significantly increase the complexity of maintaining the reliability and stability of interdependent upstream projects or branches within the OpenStack ecosystem. Such complexity highlights the importance of studying flakiness prorogation within OpenStack, as it can severely impact project stability and reliability.

\textbf{Frequent Repeated Builds on Identical Code.} Flakiness is characterized by non-deterministic testing outcomes. Unlike local testing, CI testing outcomes are transparent and observable by users. The OpenStack ecosystem employs the widely used Gerrit review tool for code reviews. Prior research~\citep{maipradit2023repeated} reported that developers in the OpenStack ecosystem unconstrainedly use the ``recheck'' command, a re-run technique to mitigate non-deterministic behavior. The recheck command is frequently invoked after the builds fail, with 55\% of reviews being identified from 41,868 code review. However, less than half of the cases do not change the outcomes of failing builds.

\subsection{Data Preparation}
\label{sec:dp}
To prepare the OpenStack data for analysis, we first perform data collection, then build-related comments extraction, and finally, flaky build identification. We describe each step below.

\textbf{Data Collection:} We leverage the RESTful API\footnote{\url{https://review.opendev.org/changes/?q=status:closed}} provided by the Gerrit tool to collect closed code reviews from OpenStack projects. 
Considering that logs from third-party testing in OpenStack are only retained temporarily---typically for one month~\cite{zuul3p}, and potentially for shorter durations (with default retention set to 7 days~\cite{nodepool})---we scrape the API on a daily basis to build this collection of one year.
As shown in Table~\ref{tab:statistic}, we capture 29,175 closed code reviews and 73,707 patch sets across 649 OpenStack projects from June 2023 to June 2024.
Since we aim to investigate the interactions between developers and CI bots, we also collect review comments (which include the change log
information) associated with each code review, yielding a total of 439,618 comments.


\textbf{Build-Related Comments Extraction.}
We focus on Zuul CI, the primary pipeline-based project gating system in OpenStack, which facilitates the execution of tests and automated tasks triggered by code review events. We deem Zuul representative for our study since it is the most commonly employed CI system in code review. Indeed, 691 OpenStack projects~\cite{zuul} of 1,336 projects~\cite{openstack} (52\%) use Zuul CI.
We extract build-related comments by matching patterns (\path{[merge|build] [succeeded|failed]}) in the review comments posted by the Zuul bot. This approach enables us to identify 139,768 comments reporting build outcomes.


\textbf{Flaky Build Identification.}
From the 139,768 collected Zuul comments, we select pairs of build invocations, which are repeated for the same patch set and have successful and failing outcomes. We classify the failing builds as flaky builds. 
The rightmost portion of Figure~\ref{fig:exm} labelled \circled{FF0000}{C} illustrates an instance of detecting flaky builds in code reviews. Here, the build outcome for \path{287458[...]c6fde2} (the build invocation in the upper-right part of Figure~\ref{fig:exm}) switched from failure to success after rechecking (the build invocation in the bottom-right part of Figure~\ref{fig:exm}), thereby identifying it as a flaky build.
The regular expression that we use to extract build information from comments can be found in the replication package. Applying that \fix{pattern-matching} query to our review comment corpus, we detect 29,911 flaky builds
across 1,651 unique Zuul jobs spanning 400 OpenStack projects.

\begin{table}[t]
\caption{Summary statistics of collected data sets.}
    \label{tab:statistic}
    \centering
    \begin{tabular}{lr}
    \toprule
    \# Studied Projects & 649 \\
Studied Period & June 2023 -- June 2024 \\
\# Code reviews & 29,175 \\
\# Review Comments (change log) & 439,618 \\
\# Patch sets & 73,707 \\
Avg. - Med. - Max. \# patch sets per CR & 2.5 - 1 - 450 \\
\# Zuul comments & 139,768 \\
\# Flaky jobs & 1,651 \\
\# Flaky builds & 29,911 \\
\bottomrule
    \end{tabular}
\end{table}

\section{Frequency of Cross-Project Flakiness (RQ1)}
\label{sec:rq1}
In this RQ, we investigate the extent to which cross-project flakiness occurs and the manner in which it is dispersed across the ecosystem.

\subsection{Approach}

To identify cross-project flakiness at a coarse-grained level (i.e., flaky builds), we examine whether a Zuul job exhibits non-deterministic behavior across multiple projects. For instance, the Zuul job ``tempest-integrated-storage'', shown in the upper-right of Figure~\ref{fig:exm}, demonstrates flaky behavior in several projects, including \path{Cinder}, \path{Glance}, and \path{Zaqar}. Among the 29,911 flaky builds identified across 1,651 unique Zuul jobs, we categorize each job based on whether its flakiness recurs in the projects it was executed. 

\fix{For comprehensive coverage in the OpenStack ecosystem, we aggregate builds from all 1,651 flaky Zuul jobs to analyze different levels of flakiness. Through the use of the Zuul API,\footnote{\url{https://zuul.opendev.org/api/tenant/openstack/builds}} we collect additional build data, ensuring each flaky Zuul job included a sufficient volume of builds (specifically, at least 200) to accurately represent its occurrence and impact within the OpenStack code reviews. We subsequently identify whether a Zuul job has been executed in multiple projects.}

To examine the frequency of cross-project flakiness, we monitor instances of flakiness by analyzing CI bot comments that indicate flaky jobs (e.g., \circled{FF0000}{C} in Figure~\ref{fig:exm}) within code reviews. This approach enables us to timestamp the occurrences of flakiness and trace their propagation over time. We identify whether flaky builds occurred across projects based on the data collected at that time for propagation analysis. 

\subsection{Results}
\minor{In this paper, we present results with respect to observations, which are followed by the evidence that supports them.}

\begin{figure}[t]
    \centering
    \includegraphics[width=\columnwidth]{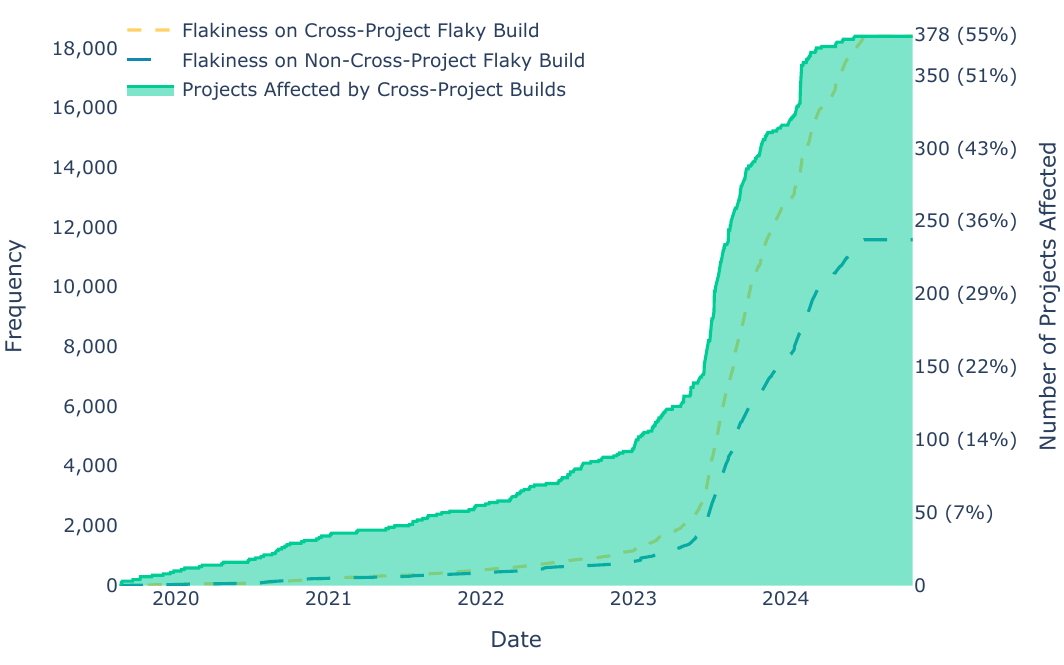}
    \caption{Propagation of flakiness over time.}
    \label{fig:pro}
\end{figure}

\textbf{\minor{\textit{Observation 1.}} \ul{Cross-project flakiness affects half of OpenStack projects and is more frequent than non-cross-project flakiness.}}
Figure~\ref{fig:pro} shows that a total of 378 OpenStack projects, approximately 55\% of all projects (691 projects that employ Zuul CI~\cite{zuul}), have been affected by cross-project flakiness. This high proportion of impacted projects suggests that flakiness is not merely an isolated issue, but instead permeates across the majority of OpenStack projects, affecting collaborative efforts and shared resources.

We find that cross-project flakiness not only affects many projects but also occurs more frequently than non-cross-project flakiness. In total, we identify 371 unique flaky Zuul jobs that had non-deterministic behavior across multiple OpenStack projects, compared to 1,280 flaky jobs limited to a single project. Although cross-project flaky jobs (18,353 flaky builds) represent only 29\% of the non-cross-project flaky jobs (11,558 flaky builds), they contribute to nearly twice as many flaky builds. This indicates that cross-project builds are more prone to repeated failures, likely due to the shared dependencies and environmental conditions across projects. We observe that 44\% of the cross-project flaky jobs had propagated to more than two projects, while the rest of the cross-project flaky jobs propagated to two projects.


\textbf{\minor{\textit{Observation 2.}} \ul{Cross-project flakiness has shown a consistent increase over time and grows faster than non-cross-project flakiness.}}
Over the observed time frame, the occurrence of cross-project flakiness has steadily increased, we find 52 projects where cross-project flakiness occurred in 2020; however, this number increased to 378 in 2024, a sevenfold increase after four years. This rising trend implies that as OpenStack projects increasingly rely on shared Zuul jobs, the probability of flakiness propagating across projects also rises, likely due to dependencies and shared testing infrastructure in OpenStack.

Figure~\ref{fig:pro} shows that cross-project flakiness demonstrates a faster growth rate over time compared to those not. This trend highlights the increasing complexity of managing CI workflows as projects evolve to use shared resources more extensively. The frequent occurrence of cross-project flakiness also emphasizes the need for robust cross-project coordination to address these issues effectively, as unmanaged flakiness in shared jobs can lead to considerable delays in code reviews and inefficiencies across dependent projects.

\begin{tcolorbox}[colframe=black,colback=gray!40]
Cross-project flakiness affects 55\% of OpenStack projects and is implicated in twice as many flaky
builds as those not. Cross-project flakiness has grown more prevalent over time, likely due to shared dependencies and environmental inconsistencies across projects.
\end{tcolorbox}

\section{Relationship Between Test Scopes and Flakiness Levels (RQ2)}
\label{sec:rq2}
In this RQ, we investigate the influence of test scopes (e.g., unit and full-stack tests) on the dispersion of flakiness.

\subsection{Approach}

\textbf{Test Result Collection.}
For each flaky build identified in Section~\ref{sec:rq1}, we retrieve the associated build log URL, containing a unique identifier (a.k.a., UUID) to access specific job outputs (e.g., \path{job-output.txt}) and test results (e.g., \path{testr_results.html}). 

Our access to build logs is constrained by the log retention policy, which varies among the studied projects. Out of a total of 29,911 flaky builds, 13,140 were invoked during our retention period of one year. Despite these constraints, we successfully retrieve \path{testr_results.html} files from 3,369 flaky builds, spanning 370 unique jobs. We filter out data from projects that no longer exist at the time of data collection in two cases: (i) unavailable job logs, and (ii) job records no longer existing (e.g., \path{heat-functional-yoga}).

\textbf{Flaky Test Identification.}
From successfully collected test results, we identify specific test cases that had failed as flaky tests. Each flaky test is \fix{identified by the combination of its fully-qualified class and test method names, e.g., 
\path{tempest.api[...]RescueTest} and
\path{test_stable[...]volume_attached[...]} in Figure~\ref{fig:exm}, respectively. The first segment (e.g., \path{tempest}) in the class name indicates the root package name. Together with the test method name, it ensures tests with this identifier correspond to semantically equivalent code and prevent naming collisions across OpenStack projects. This identifier is also used by the test runner (\path{stestr})~\cite{unittest40:online,stestrus44:online} for reporting results and OpenSearch~\cite{opensearch}, adopted by the studied community for searching OpenStack testing and building logs.}

In total, this process yields 11,506 flaky tests, which span 57,124 flaky instances (executions). 

\begin{table*}[t]
\centering
\caption{Distribution of flaky tests and instances by scope and flakiness level in OpenStack projects.}
\label{tab:flaky_distribution}
\begin{tabularx}{\linewidth}{lp{20mm}Xrrr}
\toprule
 & \textbf{Category} & \textbf{Definition} & \textbf{\# Tests} & \textbf{\# Instances} & \textbf{Avg. Executions} \\
\midrule
\multirow{12}{*}{\rotatebox{90}{Test Scope}} & Unit & Tests individual units of code in isolation, ensuring correctness during the development phase, involving the term ``\textit{unit}'' in its identifier. & 7,202 & 30,550 & \fix{2,845}\\
\cline{2-6}
 & API & Evaluates the functionality, performance, and
security of APIs to ensure compliance with specifications,
containing ``\textit{api}'' in its identifier. & 1,285 & 10,599 & \fix{3,617} \\
\cline{2-6}
 & Resource Management & Manages resources and configurations required before all tests or releases resources after all tests in a test class. These tests typically start
with ``\textit{setUpClass}'' or ``\textit{tearDownClass}''. &636 & 8,414 & \fix{37} \\
\cline{2-6}
 & Scenario & Simulates end-to-end functionality to validate integration points between OpenStack services, containing ``\textit{scenario}'' in its identifier. & 383 & 3,065 & \fix{1,328} \\
 \cline{2-6}
 & Functional & Assesses broader functionalities of an
OpenStack project, verifying correct outputs for various
inputs, including ``\textit{functional}'' in its identifier. & 1,230 & 1,963 & \fix{488} \\
\cline{2-6}
 & Full Stack & Verifies interactions between OpenStack
components to ensure they operate correctly in combination, including ``\textit{fullstack}'' in its identifier. & 85 & 293 & \fix{885}\\
\cline{2-6}
 & Undefined & Any test that does not fit the above categories. & 685 & 2,240 & \fix{639} \\
\midrule
\multirow{8}{*}{\rotatebox{90}{Flakiness Level}} & Shared-Partial Trigger & Flakiness appears across multiple projects, with flaky tests triggered only by some of the affected projects. & 703 & 16,976 & \fix{4,453}\\
\cline{2-6}
 & Uniform Multi-Project & Flakiness spans multiple projects, with the same set of projects consistently triggering these flaky tests. & 832 & 3,949 & \fix{333} \\
 \cline{2-6}
 & Cross-Trigger & Flakiness is present in a single
project, yet its test is also triggered by other projects. & 402 & 1,926 & \fix{1,955} \\
\cline{2-6}
 & Project-Isolated & Flakiness is confined to a
single project, with the associated test triggered only
within that project. & 371 & 1,714 & \fix{162} \\
\cline{2-6}
 & Failed to obtain & The build logs are not available. & 95 & 490 & \fix{-} \\
\bottomrule
\end{tabularx}
\end{table*}

\textbf{Test Scope Classification.}
To understand the nature of these flaky tests, we systematically identify different scopes of 11,506 flaky tests. \fix{Our approach extracts all segments except the first one (e.g., \path{tempest}) from the dot-delimited fully-qualified class name for keyword-matching. The test scopes and the corresponding keywords for each test scope are built based on the OpenStack general~\cite{opendoc} and project-specific~\cite{opendoc-unit,opendoc-fullstack} documentation to reflect the testing design in the OpenStack ecosystem and avoid matching ambiguous identifiers. For example, the Neutron documentation~\cite{opendoc-unit} explains that ``Unit test modules should have the same path under neutron/tests/unit/ as the module they target has under neutron/.'' These test scopes and keywords (available in the top portion of Table~\ref{tab:flaky_distribution}) are iteratively and manually refined by the authors to minimize undefined cases (e.g., ``setUpClass''). For multiple segments matching, we prioritize classification from longer, less common test scopes (Full Stack) over shorter, more common scopes (API and Unit).}








\textbf{Flakiness Level Classification.}
To classify flakiness levels, we identify how flakiness behaved across different OpenStack projects. We categorized each flaky test into one of four levels \fix{based on the definitions} in the bottom portion of Table~\ref{tab:flaky_distribution}.

To automate flakiness categorization, we collect \fix{test results from builds obtained in RQ1. For example, we retrieve the test results of the \path{tempest-integrated-storage} job, shown in the upper-right of Figure~\ref{fig:exm}. In this example, three OpenStack projects (\path{Cinder}, \path{Glance}, and \path{Zaqar}) executed the flaky test identified in this job. If this flaky test only appears flaky in \path{Glance}, we categorize it as \textit{Cross-Trigger} flakiness since the flakiness is limited to one project. When the flakiness occurred in both \path{Cinder} and \path{Glance}, we classify these as \textit{Shared-Partial Trigger} flakiness since the flakiness propagated more than one project, but not all projects run it. If the flakiness was observed in all three projects, we classify this as \textit{Uniform Multi-Project} flakiness since the flakiness propagated across all projects. Conversely, if flakiness was only observed in \path{Glance} and the test was exclusively run by \path{Glance}, this is classified as \textit{Project-Isolated} flakiness.}

For the two phenomena we discovered, cross-project flakiness is a combination of \textit{Shared-Partial Trigger} and \textit{Uniform Multi-Project} flakiness, while inconsistent flakiness results from a combination of \textit{Cross-Trigger} and \textit{Shared-Partial Trigger} flakiness.

\subsection{Results}
We plot the relationship between test scopes and levels of flakiness by using the parallel sets~\cite{kosara2006parallel}. Parallel sets
are variants of parallel coordinates in which the width of lines that connect sets corresponds to the frequency of their co-occurrence. To plot more accurate results in identifying high-impact flakiness within OpenStack, we \textbf{exclude flaky tests that appeared in only a single patch set}. This filtering yields 2,403 flaky tests from 11,506 flaky instances, each occurring in multiple patch sets.



\textbf{\minor{\textit{Observation 3.}} \ul{Unit tests are the most likely to exhibit flakiness.}}
The top portion of Table~\ref{tab:flaky_distribution} illustrates the distribution of flaky tests and their instances across various test scopes in OpenStack projects. Unit tests account for the largest amount of flaky tests (7,202) and instances (30,550). According to the OpenStack documentation~\cite{opendoc}: ``Most projects have a large number of unit tests and expect to have near-complete code coverage with those tests.'' Thus, it is not unexpected that the large quantity of unit tests overall accounts for a large number of flaky tests. In our data set, we find the most commonly flaky unit test~\cite{unittest1}, exhibiting flakiness 72 times in the Nova project. The test verifies that the database schema after migrations is in sync with the expected models in the code. 

API tests are the second most frequently implicated, with 1,285 flaky tests and 10,599 instances. Flakiness in API tests typically stems from their reliance on external components and third-party services, which may not always be stable or available. Furthermore, the intrinsic complexity of cloud environments, characterized by interacting components operating concurrently and often asynchronously, exacerbates this issue. Such environments can induce non-deterministic behavior, particularly when interactions are not managed or simulated within test scenarios~\cite{musavi2016experience}. We find that the API test~\cite{apitest1} that is most frequently flaky impacts 14 OpenStack projects. The test evaluates the functionality of uploading a volume to create an image using the API.

A substantial proportion of flaky instances are associated with resource management tests. These tests orchestrate resources under conditions shared across multiple tests within the same test class, ensuring a consistent starting environment and proper cleanup after tests to prevent side effects on subsequent tests. Flakiness in these tests is likely due to concurrency issues or dependencies arising from the shared state or resources. 

\fix{To reduce the influence of flaky test execution frequencies on our observations, we estimate these frequencies across OpenStack by computing the average number of executions for each test scope in our collected test results, as shown in Table~\ref{tab:flaky_distribution}. While we observe more flaky unit tests in our data set, their average execution frequency (2,845) is lower than that of API tests (3,617), which further supports that Observation 3 is not caused by frequent executions of unit tests in CI.}


\textbf{\minor{\textit{Observation 4.}} \ul{Flakiness in cross-project tests is more prevalent than in single-project tests.}}
The bottom portion of Table~\ref{tab:flaky_distribution} presents the distribution of flakiness levels, showing that \textit{Shared-Partial Trigger} and \textit{Uniform Multi-Project} flakiness are the most prevalent, accounting for 16,976 and 3,949 instances, respectively. This trend indicates the spread of flakiness affecting tests executed across multiple projects, accounting for 1,535 ($703+832$) tests under cross-project flakiness phenomenon. The \textit{Cross-Trigger} flakiness is detected in 402 tests. Similar to \textit{Shared-Partial Trigger} flakiness, they belong to the inconsistent flakiness phenomenon, which affects a total of 1,105 ($703+402$) tests.
In Section~\ref{sec:rq3}, we further examine why specific flakiness types predominantly impact one or a few projects. Conversely,
\textit{Project-Isolated} and \textit{Uniform Multi-Project} flakiness occur under ideal circumstances, either affecting only one project or all projects uniformly.

\begin{figure}[t]
    \centering
    \includegraphics[width=.5\textwidth]{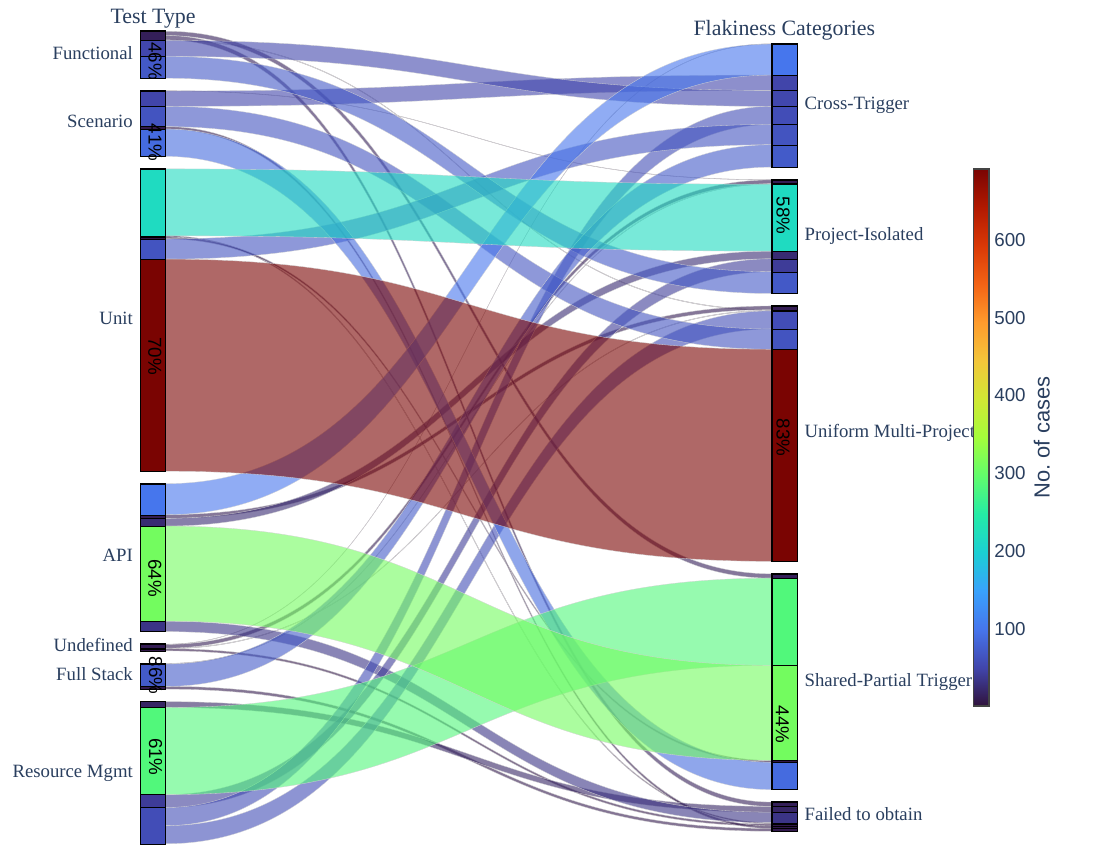}
    \caption{Parallel sets between test scopes and flakiness levels in OpenStack projects.}
    \label{fig:pic4}
\end{figure}

\textbf{\minor{\textit{Observation 5.}} \ul{Beyond functional and full-stack tests, test scopes demonstrate correlations with cross-project flakiness, as unit tests, intended to validate individual components, exhibit flakiness across multiple projects.}} In Figure~\ref{fig:pic4}, functional and full-stack tests are strongly associated with \textit{Project-Isolated} and \textit{Cross-Trigger} flakiness, accounting for 46\% and 86\% of cases, respectively. Flakiness in functional tests frequently remains confined to a single project. As stated in OpenStack's documentation~\cite{opendoc}, ``Functional tests validate a project in operation (e.g., if the project is an API server, the server will actually be running), though interactions with \textbf{external components are kept to a minimum}.'' 

\fix{A substantial percentage (70\%) of unit tests exhibits \textit{Uniform Multi-Project} flakiness, indicating that for all OpenStack projects where these unit tests were executed, the flakiness occurred consistently. For example, \path{test_servers}~\cite{test_servers} test, used in projects such as \path{Glance} and \path{Cinder}, has spread flakiness due to its reliance on \path{OpenStackSDK} to interface with the Nova API. By contrast, API tests (64\%) and scenario tests (41\%) show 
\textit{Shared-Partial Trigger} flakiness, where flaky behavior appears in multiple projects but only within a subset of them. This is expected, as they gate commit integrations into OpenStack, which may behave differently based on environmental contexts. Similarly, 61\% of resource management tests follow this pattern, suggesting possible misconfigurations or poor resource management.}


\begin{tcolorbox}[colframe=black,colback=gray!40]
About half of API tests and scenario tests are flaky across projects, likely due to similarity to integration-level testing. Substantial unit tests are evaluated with flakiness in multiple projects, indicating potential misalignments between the test design and its intended scope. More than half of resource management tests show inconsistent flakiness, suggesting misconfigurations or issues with resource management within the test design.
\end{tcolorbox}

\section{Causes of Inconsistent Flakiness (RQ3)}
\label{sec:rq3}
In this RQ, we qualitatively investigate why flakiness occurs in specific OpenStack projects rather than affecting all projects that execute the same flaky test. \fix{We analyze the root causes of inconsistent flakiness instead of cross-project flakiness to provide a control condition, since the test logic of these inconsistently flaky tests is observed to be stable in the ``control'' project, the non-deterministic behavior could be attributed to possible project-specific configurations rather than intrinsic non-deterministic behavior, which has been widely analyzed in the single-project context~\cite{eck2019understanding}.}

\subsection{Approach}
We employ an open coding approach~\cite{charmaz2014constructing} to classify \fix{inconsistently} flaky tests. Open coding, a qualitative analysis method, involves classifying the artifacts under investigation (flaky tests in this case) according to emergent concepts. Following this initial coding, we apply open card sorting~\cite{opencard} to organize low-level codes into higher-level categories. Below, we detail our coding saturation, coding strategy, and card-sorting procedures.

\textbf{Coding Saturation.} This RQ focuses on 1,105 \fix{inconsistently flaky tests} spanning \fix{18,902 ($16,976+1,926$)} instances \fix{or test executions (a flaky test can manifest its flakiness behavior multiple times)}; specifically, \textit{Cross-Trigger} and \textit{Shared-Partial Trigger} flakiness manifests in a subset of projects but not all of them. Given the impracticality of coding all 1,105 tests \fix{for over ten thousand instances}, we code a \fix{subset of these flaky tests} to enable analytic generalization and reach theoretical saturation~\cite{eisenhardt1989building} \fix{for exploring the underlying cause of this novel phenomenon in the OpenStack ecosystem}.

To discover as complete of a list of reasons for inconsistent flakiness as possible, we strive for theoretical saturation~\cite{eisenhardt1989building} to achieve analytic generalization. We set our saturation criterion to 10: the first two authors continue to code randomly selected flaky tests until no new codes have been discovered for 10 consecutive flaky tests. Since a flaky test could manifest many flaky instances, we adopt a relatively small saturation criterion. Furthermore, a single flaky test may have multiple labels.

We produce an initial set of codes after the first 30 flaky tests. Based on this set of codes, the first two authors independently code another 30 flaky tests and then calculate Fleiss's kappa agreement~\cite{fleiss1971measuring} of this iteration among all raters. The kappa agreement of this iteration is 0.726 or ``Substantial'' agreement~\cite{viera2005understanding}.
Based on this encouraging result, the remaining data is then coded by the first author. Finally, we reach saturation after coding \fix{in total of 110} inconsistently flaky tests, spanning \fix{671 test executions}. Since codes that emerge late in the process may apply to earlier reviews, we perform another pass over all of the flaky tests to correct miscoded entries and test the level of agreement of our constructed codes. 
\fix{These 110 coded flaky tests span all test scopes defined in RQ2 and account for 62\% of projects exhibiting inconsistent flakiness.}

\textbf{Coding Strategy.} Our coding strategy aligns with OpenStack's official documentation~\cite{opendoc-how} on handling test failures. First, we review available CI logs and error messages to identify symptoms of flaky tests. Given that flaky test causes can be non-deterministic, we also examine relevant inline comments in the code reviews. Finally, we search related bug reports based on the test symptoms to identify possible causes. \fix{This process is extensively conducted on triangulating CI logs, error messages, and bug reports for 671 test executions.}

\textbf{Card Sorting.} We apply open card sorting to construct a taxonomy of factors that contribute to the occurrence of flakiness in one software project but not in others. 
Open card sorting produces generate general categories from our low-level codes. The open card sorting includes two steps. First, the coded cases are merged into cohesive groups that can be represented by a similar subcategory. Second, the related subcategories are merged to form categories that can be summarized by a short title.

\subsection{Results}
\textbf{\minor{\textit{Observation 6.}} \ul{Inconsistent flakiness across projects is primarily caused by event-related issues, such as server downtime or temporarily unavailable resources, but we also observe dependency issues of the patch, and inconsistent build configurations.}} In Table~\ref{tab:reason}, \textit{Event-related flakiness} is the most frequently occurring reason for inconsistent flakiness. These inconsistencies can stem from race conditions in CI or dependency issues from the patch. Interestingly, a small number of inconsistencies were linked to build configurations. Only one specific flaky test occurred due to an unordered collection comparison in the \verb|assertEqual| function.

\begin{table}[t]
\centering
\caption{Definition and frequency of reasons for inconsistent flakiness.}
\label{tab:reason}
\begin{tabularx}{\columnwidth}{p{30mm}rX}
\toprule
\textbf{Reason}  & \textbf{\#} & \textbf{Definition} \\
\midrule
\cellcolor{grayrow}
\textbf{Event-related flakiness}& \cellcolor{grayrow}\textbf{\fix{98}}  & \multirow{3}{\linewidth}{Flakiness in this category is caused by external or transient events impacting the stability of tests across different projects. These events (race conditions \minor{in CI}) may be unpredictable and are often resolved in the short term.}   \\
\cline{1-2}
 Backend server down & \fix{25} &   \\
 \cline{1-2}
 Temporarily unavailable resource & \fix{21}&   \\
 \cline{1-2}
 Short-term job fixes & \fix{17} &  \\
 \cline{1-2}
 Networking issues	& \fix{14}&  \\
  \cline{1-2}
 Unstable job within period	& 13& \\
  
  \cline{1-2}
 Simultaneous reading conflicts & \fix{8}&  \\
 \midrule
 \cellcolor{grayrow}
 \textbf{Dependency-related flakiness} & \cellcolor{grayrow}\textbf{\fix{23}} & \multirow{3}{\linewidth}{Flakiness caused by dependencies on external components or project transitions, affecting tests differently across projects due to versioning, updates, or reliance on specific branches.}  \\
 \cline{1-2}
Bump issues	& \fix{17}&  \\
\cline{1-2}
Dependent issues &2 &	 \\
\cline{1-2}
Supporting branch transition & 2& \\
\cline{1-2}
Dependent with test change & 2&  \\
\midrule
 \cellcolor{grayrow}
 \textbf{Configuration-related flakiness} & \cellcolor{grayrow} \textbf{\fix{23}}& \multirow{3}{\linewidth}{This type of flakiness is due to configuration issues within the testing environment or infrastructure, leading to inconsistencies in test outcomes across projects.}  \\
 \cline{1-2}
Slow deprovisioning of virtual machines & \fix{19}&	 \\
\cline{1-2}
Temporary server disabling & 4&	 \\
\midrule
\cellcolor{grayrow}
 \textbf{Other} & \cellcolor{grayrow}\textbf{1} & An outlier category for flakiness caused by a miscellaneous or unordered collection issue not fitting the main categories above.   \\
\bottomrule
\end{tabularx}
\end{table}

\begin{figure}[t]
    \centering
    \includegraphics[width=\columnwidth]{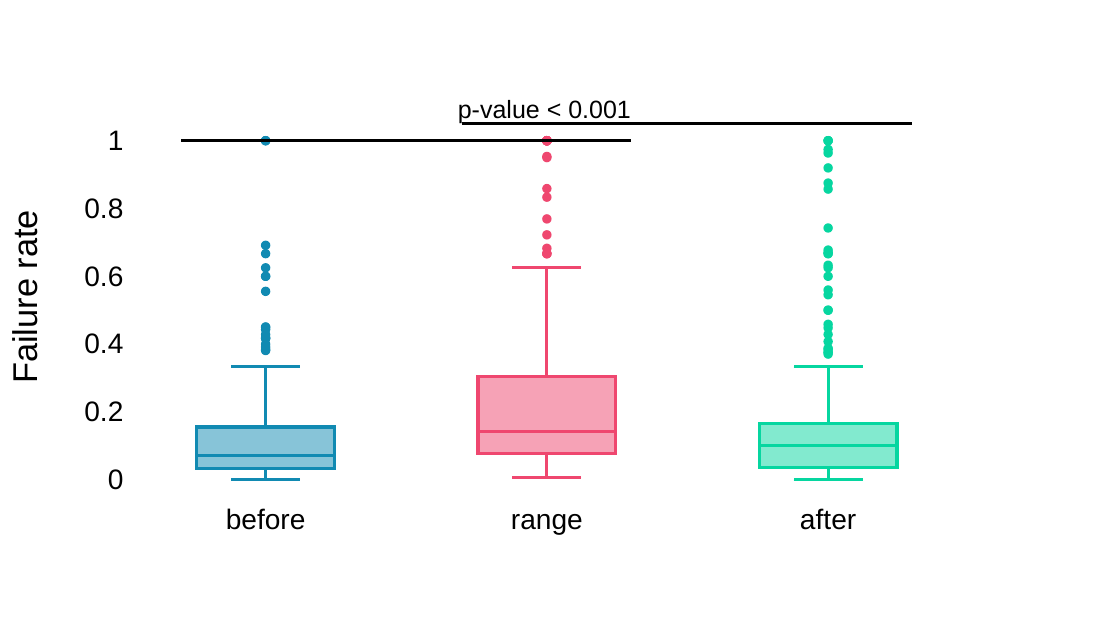}
    \caption{Failure rates of each Zuul job before, within, and after flaky range for inconsistent flakiness.}
    \label{fig:timing}
\end{figure}

In total, \fix{98} cases exhibit inconsistent flakiness primarily due to event-related causes. For instance, flakiness was observed when the backend server was down at a particular time in an API test. A related bug report~\cite{bug_1833757} stated:

\begin{tcolorbox}[colback=gray!5!white, arc=0pt, outer arc=0pt, colframe=gray!30!black, boxsep=5pt, boxrule=0.5pt,size=fbox,breakable]\footnotesize
I looked at the logs from link in comment \#1 for the request to create volume [...]. The request was serviced by the cinder-scheduler running on controller-2, and the logs show the RBD backend was down at that time.
\end{tcolorbox}

Another example of event-related flakiness occurred in a scenario test due to temporarily unavailable resources. The error log indicated:

\begin{tcolorbox}[colback=gray!5!white, arc=0pt, outer arc=0pt, colframe=gray!30!black, boxsep=5pt, boxrule=0.5pt,size=fbox,breakable]\footnotesize
Traceback (most recent call last):
  File ``[...]'', line 521, in [...]
BlockingIOError: [Errno 11] Resource temporarily unavailable
\end{tcolorbox}

This type of flakiness tends to occur repeatedly within a short timeframe. We then investigate whether the failure rates of the flaky jobs for these flaky tests differed significantly from the flakiness possibly caused by race conditions \minor{in CI}. We define the flaky range as the duration from the first identified instance of flakiness to the latest recorded instance in our data set for each flaky job. This range provides perspective on the persistence of flaky behavior over time and across projects. We then examine differences in failure rates within, before, and after the flaky range of each job by performing a Mann-Whitney test~\cite{mann1947} (two-tailed, unpaired, $\alpha = 0.05$). To measure the effect size, we also apply Cliff's $\delta$~\cite{cliff1993dominance}, a non-parametric effect size measure. Effect size is characterized as \emph{negligible} if $\delta < 0.147$, \emph{small} if $0.147 \leq |\delta| <0.33$, \emph{medium} if $0.33 \leq |\delta| <0.474$, \emph{large} if $0.474 \leq | \delta|$~\cite{romano:2006}.


\textbf{\minor{\textit{Observation 7.}} \minor{\ul{Inconsistent flakiness caused by race conditions in CI has significantly increased the failure rates.}}}
Figure~\ref{fig:timing} indicates that failure rates of flaky jobs within the flaky range are significantly higher than before and after the flaky range. The Cliff's $\delta$ of the before and within the flaky range is \emph{medium}. We find a \emph{small} effect between within and after the flaky range.
This confirms our qualitative exploration, where the principal cause happened in a short timeframe.

\minor{Beyond event-related issues}, \fix{23} cases exhibit inconsistencies related to dependency issues. A notable example involves a version bump issue, where developers rechecked the patch (triggering the failed job to execute again on the Gerrit platform) with an inline comment for a flaky full-stack test:
\begin{tcolorbox}[colback=gray!5!white, arc=0pt, outer arc=0pt, colframe=gray!30!black, boxsep=5pt, boxrule=0.5pt,size=fbox,breakable]\footnotesize
recheck requirements bump merged
\end{tcolorbox}

Inconsistently flaky tests may also be caused by the temporary disabling of test cases when a dependent project transitions to an unmaintained branch~\cite{unmaintained}. For instance, a developer rechecked the patch set with the following comment to another review:
\begin{tcolorbox}[colback=gray!5!white, arc=0pt, outer arc=0pt, colframe=gray!30!black, boxsep=5pt, boxrule=0.5pt,size=fbox,breakable]\footnotesize
recheck - [...] 906582 has merged
\[\Downarrow Review~\#906582's~Description\]
{[requirements]} Transition Yoga to Unmaintained. 
With the new TC resolution the community replaces Extended Maintenance with Unmaintained status. So this patch transitions stable/yoga branch to Unmaintained [...]

\end{tcolorbox}

We \minor{also} find flakiness can result from differing CI configurations (23). For example, one developer rechecked the patch:
\begin{tcolorbox}[colback=gray!5!white, arc=0pt, outer arc=0pt, colframe=gray!30!black, boxsep=5pt, boxrule=0.5pt,size=fbox,breakable]\footnotesize
recheck Fixed IP address [...] is already in use on instance likely caused by slow vm
\end{tcolorbox}

\noindent In such cases, the resource is likely occupied by other processes, as suggested by the error message:
\begin{tcolorbox}[colback=gray!5!white, arc=0pt, outer arc=0pt, colframe=gray!30!black, boxsep=5pt, boxrule=0.5pt,size=fbox,breakable]\footnotesize
Target share type is still in use. 
\end{tcolorbox}

Solutions for these issues include limiting thread concurrency in CI environments~\cite{review920766} or increasing the swap memory size~\cite{review921420}. Another potential solution is silencing services in the CI configuration:
\begin{tcolorbox}[colback=gray!5!white, arc=0pt, outer arc=0pt, colframe=gray!30!black, boxsep=5pt, boxrule=0.5pt,size=fbox,breakable]\footnotesize
recheck [...] 922902 was merged
\[\Downarrow Review~\#922902~Description\]
Disable openstack-cli-server. The server does not support switching credentials well and causes problems with resources created for [...] tests.
\end{tcolorbox}

\begin{tcolorbox}[colframe=black,,colback=gray!40]
Results in RQ3 suggest a need to standardize CI configuration as well as dependency management across the ecosystem. The primary cause of inconsistent flakiness is unanticipated event-related issues like server downtime. Inconsistent flakiness can also arise from dependency-related issues and configuration issues in CI setups, indicating that CI setting variations can also lead to flakiness.
\end{tcolorbox}

\section{\fix{Time Wasted Due to Cross-Project Flakiness}}
\label{sec:time}
In Section~\ref{sec:rq1}, we find that cross-project flakiness is a prevalent phenomenon at a coarse-grained level. To further measure the subsequent impact of cross-project flakiness, in this section, we study the time it wastes during code review. Specifically, our goal is to compare the waste caused by cross-project flakiness to that of flakiness occurring within a single project. In part \circled{FF0000}{C} of Figure~\ref{fig:exm}, flakiness is also associated with slower code review decisions and the consumption of additional build resources. In total, these add a delay of 2h 07m 05s. To more broadly quantify the impact of cross-project flakiness, we measure this additional build execution time across the 29,911 identified flaky builds. We perform Mann-Whitney tests to assess the significance of the difference between cross-project and non-cross-project flaky jobs.

\begin{figure}[t]
    \centering
    \includegraphics[width=\columnwidth]{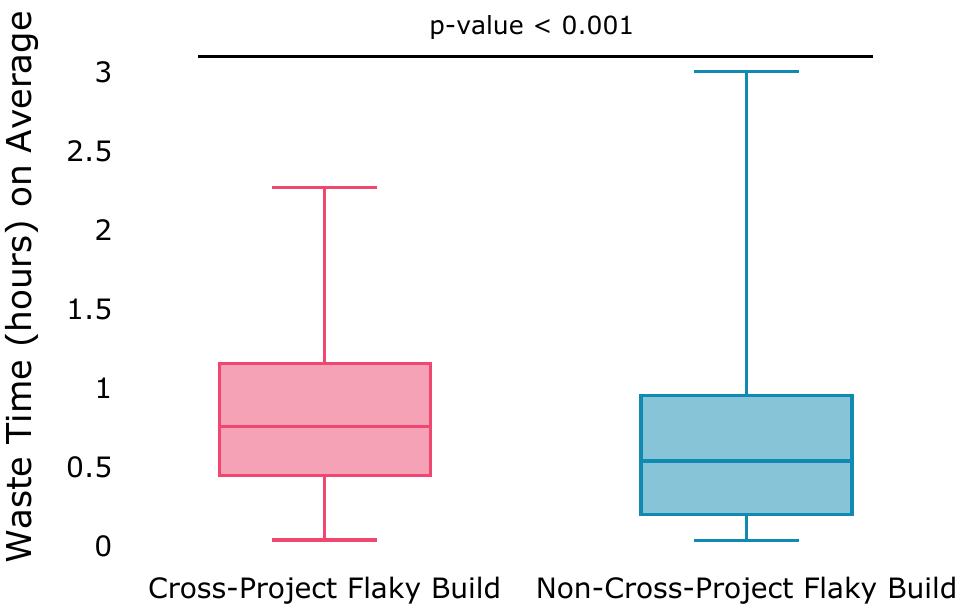}
    \caption{Time wasted (hours) of each Zuul job.}
    \label{fig:diff}
\end{figure}

\textbf{\minor{\textit{Observation 8.}} \ul{Cross-project flakiness generates more time waste than flakiness limited to a single project.}} 
We find that flakiness has cumulatively wasted 1,156 days on code reviews, averaging 0.93 hours of additional time per flaky build. Figure~\ref{fig:diff} shows that cross-project flaky builds have a higher median of time waste than non-cross-project flaky builds. The Mann-Whitney test confirms this with a significant difference ($p < 0.001$). The effect size for this significance of the difference is \emph{small}, i.e., $0.147 \leq |\delta| <0.33$. It indicates that cross-project flakiness is associated with slower completion of code reviews (due to the re-execution of unreliable CI jobs) than flaky tests that occur within a single project. This also suggests that general flakiness prolongs code review cycles in OpenStack. If flakiness propagates across multiple projects, it produces significantly more time and resource waste.

\section{Developer Feedback}
To gain deeper insights into cross-project and inconsistent flakiness,
we conduct a developer questionnaire\footnote{The questionnaire was approved by the Kyushu University, Japan ethics board (2024-26) \fix{and is available at \url{https://tinyurl.com/croflakiness}.}} to understand how developers perceive these issues. Our goals are to: (i) gain perspective on the practitioner experience with these challenges and how these problems are typically addressed, (ii) gather feedback on the causes of inconsistent flakiness, and (iii) explore existing tool support and opportunities for developing new tool support. 
We distributed an open invitation to participate in our questionnaire via the OpenStack discussion mailing list and reached out to 833 developers who had participated in the studied code reviews. We accepted questionnaire responses for one month and received valid responses from 15 developers. 

\begin{table}[t]
    \centering
        \caption{Respondents' role and their experiences in OpenStack as well as with cross-project and inconsistent flakiness.}
    \label{tab:demo}
    \resizebox{\columnwidth}{!}{\begin{tabular}{lllll}
    \toprule
         \multirow{2}{*}{\textbf{ID}} & \multirow{2}{*}{\textbf{Role}} & \multicolumn{3}{c}{\textbf{Experience}} \\ 
         & & OpenStack & Cross-project & Inconsistent  \\
         \midrule
R$_{1}$ & Contributor & 3-5 years & No & Yes \\ 
R$_{2}$ & Core developer & More than 5 years & Yes & Yes \\ 
R$_{3}$ & Core developer & More than 5 years & Yes & Yes \\ 
R$_{4}$ & Contributor & More than 5 years & No & Yes \\ 
R$_{5}$ & Contributor & More than 5 years & Yes & No \\ 
R$_{6}$ & Core developer & 3-5 years & Yes & Yes \\ 
R$_{7}$ & Contributor & More than 5 years & No & Yes \\ 
R$_{8}$ & Project maintainer & More than 5 years & Yes & Yes \\ 
R$_{9}$ & Contributor & More than 5 years & Yes & No \\ 
R$_{10}$ & Contributor & More than 5 years & Yes & Yes \\ 
R$_{11}$ & Core developer & More than 5 years & Yes & Yes \\ 
R$_{12}$ & Consultant for contributors & 1-2 years & No & Yes \\ 
R$_{13}$ & Contributor & 1-2 years & No & Yes \\ 
R$_{14}$ & Contributor & More than 5 years & Yes & Yes \\
R$_{15}$ & Contributor & 1-2 years & Yes & Yes \\ \bottomrule
    \end{tabular}}
\end{table}

Table~\ref{tab:demo} provides an overview
of the demographics of our respondents. Among the 15 respondents ($R_{1}$--$_{15}$), ten reported having over five years of experience contributing to OpenStack projects. 
Ten participants are contributors or consultants for contributors, while five are core developers or project maintainers.
We analyze open-ended questionnaire responses using a card-sorting method. 
Below, we summarize the key findings, with the full questionnaire design available in our online appendix.

\textbf{Cross-Project Flakiness.}
Ten respondents indicated that they had encountered cross-project flakiness in OpenStack. In terms of addressing cross-project flaky tests differently from those in a single project, two respondents noted an increase in complexity due \fix{to the need for substantial developer collaboration across multiple teams. For example, $R_5$ explained that \enquote{Less likely to engage due to higher burden (having to engage multiple teams),} whereas $R_{10}$ emphasized the challenges of collaboration and obtaining feedback on such flakiness, particularly from the perspective of a newcomer:}
\begin{formal}
    \color{coolblack}{\small\textit{``I personally do a recheck, I am \textbf{new to contributing} and I find it \textbf{difficult to get help from the community} so I often just \textbf{ignore it or try to keep away from contributing}. I have reviews up which I \textbf{never get feedback on}, also \textbf{asking via IRC does not help} so I \textbf{do not really want to engage in fixing tests or systems}.''}}
\end{formal}

Three respondents mentioned common causes, such as race conditions, concurrency, and infrastructure-based challenges on the testing platform ($R_7$, $R_9$, and $R_{14}$). Two respondents explained that they often do not take further actions beyond re-checking and waiting for a subsequent build. Two respondents suggested verifying shared dependencies, environments, and test fixtures in different projects ($R_{12}$ and $R_{13}$).

$R_{8}$ remarked, \enquote{I prefer to view OpenStack as a \textbf{collective effort}, so failures in any one component are a failure in OpenStack as a whole.} Indeed, OpenStack provides OpenSearch, a community-maintained cluster that monitors failing tests and jobs. $R_{12}$ similarly highlighted the importance of reviewing environments and test fixtures to understand why a test is flaky. $R_{1}$ added, \enquote{Cross-project is a sign of \textbf{incorrect structure}.} 

\textbf{Inconsistent Flakiness.}
The 13 respondents reported encountering inconsistent flakiness. In terms of causes beyond those already identified in Section~\ref{sec:rq3}, $R_{12}$ described \enquote{test fixtures from different vendors running different firmware versions.} Table~\ref{tab:sur} shows that other responses provide support for issues identified in Section~\ref{sec:rq3}, such as networking issues. 

Developers expressed diverse opinions about the difficulty of debugging each cause. Event-related causes are generally perceived to be the most challenging, potentially due to that external factors can be unpredictable. $R_{12}$ also cited test fixtures from different vendors as especially difficult to debug.

\begin{table}[t]
\centering
\begin{threeparttable}

    \caption{Causes of inconsistent flakiness and their debugging difficulty from respondents.}
    \label{tab:sur}
    \begin{tabular}{lp{2cm}p{3cm}}
        \toprule
        \textbf{Causes} & \textbf{Distribution} & \textbf{Debugging difficulty} \\
        \midrule
        Event-related & 
        42.9\%
        \begin{tikzpicture}
            \begin{axis}[
                width=2.5cm, height=1.8cm, 
                xbar,
                symbolic y coords={Y},
                ytick=\empty, 
                axis lines=none,
                xmin=0, xmax=50,
                bar width=6pt,
                enlarge y limits=0.5
            ]
                \addplot[fill=gray] coordinates {(42.9,Y)};
            \end{axis}
        \end{tikzpicture} 
        & 
        \begin{tikzpicture}
            \begin{axis}[
                width=4.5cm, height=1.8cm,
                xbar stacked,
                symbolic y coords={Y},
                ytick=\empty,
                axis lines=none,
                xmin=0, xmax=15,
                bar width=6pt,
                enlarge y limits=0.5
            ]
                \addplot[fill=my4grey] coordinates {(5,Y)};
                \addplot[fill=my3grey] coordinates {(5,Y)};
                \addplot[fill=my2grey] coordinates {(3,Y)};
                \addplot[fill=my1grey] coordinates {(1,Y)};
            \end{axis}
        \end{tikzpicture} \\
        
        Dependency-related & 
        28.6\%
        \begin{tikzpicture}
            \begin{axis}[
                width=2.5cm, height=1.8cm,
                xbar,
                symbolic y coords={Y},
                ytick=\empty,
                axis lines=none,
                xmin=0, xmax=50,
                bar width=6pt,
                enlarge y limits=0.5
            ]
                \addplot[fill=gray] coordinates {(28.6,Y)};
            \end{axis}
        \end{tikzpicture} 
        & 
        \begin{tikzpicture}
            \begin{axis}[
                width=4.5cm, height=1.8cm,
                xbar stacked,
                symbolic y coords={Y},
                ytick=\empty,
                axis lines=none,
                xmin=0, xmax=15,
                bar width=6pt,
                enlarge y limits=0.5
            ]
                \addplot[fill=my4grey] coordinates {(3,Y)};
                \addplot[fill=my3grey] coordinates {(3,Y)};
                \addplot[fill=my2grey] coordinates {(5,Y)};
                \addplot[fill=my1grey] coordinates {(4,Y)};
            \end{axis}
        \end{tikzpicture} \\
        
        Configuration-related & 
        14.3\%
        \begin{tikzpicture}
            \begin{axis}[
                width=2.5cm, height=1.8cm,
                xbar,
                symbolic y coords={Y},
                ytick=\empty,
                axis lines=none,
                xmin=0, xmax=50,
                bar width=6pt,
                enlarge y limits=0.5
            ]
                \addplot[fill=gray] coordinates {(14.3,Y)};
            \end{axis}
        \end{tikzpicture} 
        & 
        \begin{tikzpicture}
            \begin{axis}[
                width=4.5cm, height=1.8cm,
                xbar stacked,
                symbolic y coords={Y},
                ytick=\empty,
                axis lines=none,
                xmin=0, xmax=15,
                bar width=6pt,
                enlarge y limits=0.5
            ]
                \addplot[fill=my4grey] coordinates {(3,Y)};
                \addplot[fill=my3grey] coordinates {(4,Y)};
                \addplot[fill=my2grey] coordinates {(5,Y)};
                \addplot[fill=my1grey] coordinates {(2,Y)};
            \end{axis}
        \end{tikzpicture} \\

        Others & 
        14.3\%
        \begin{tikzpicture}
            \begin{axis}[
                width=2.5cm, height=1.8cm,
                xbar,
                symbolic y coords={Y},
                ytick=\empty,
                axis lines=none,
                xmin=0, xmax=50,
                bar width=6pt,
                enlarge y limits=0.5
            ]
                \addplot[fill=gray] coordinates {(14.3,Y)};
            \end{axis}
        \end{tikzpicture} 
        & 
        \begin{tikzpicture}
            \begin{axis}[
                width=4.5cm, height=1.8cm,
                xbar stacked,
                symbolic y coords={Y},
                ytick=\empty,
                axis lines=none,
                xmin=0, xmax=15,
                bar width=6pt,
                enlarge y limits=0.5
            ]
                \addplot[fill=my4grey] coordinates {(3,Y)};
                \addplot[fill=my3grey] coordinates {(2,Y)};
                \addplot[fill=my2grey] coordinates {(1,Y)};
                \addplot[fill=my1grey] coordinates {(5,Y)};
            \end{axis}
        \end{tikzpicture} \\
        
        \bottomrule
    \end{tabular}
    \begin{tablenotes}
      \small 
      \item The darker bars suggest a greater difficulty level in the debugging process.
    \end{tablenotes}
  \end{threeparttable}
\end{table}

    

\textbf{Possible Tool Support.} Ten respondents believed that tooling could help mitigate inconsistent flakiness. For example, $R_{3}$ pointed out that tools can overcome this organizational or infrastructure barrier. Four other respondents have no opinion on it, while $R_{10}$ argued the opposite position, stating that \enquote{I honestly do not [agree] here, I think more tests should be [set up and] verified, but I also believe that it must not be a burden to enter the contributor community, which I currently really [feel] like [it is].} Similarly, $R_{8}$ described why they felt that prior tool-based attempts to mitigate inconsistent flakiness did not succeed:
\begin{formal}
    \color{coolblack}{\small\textit{``We've had useful tools for this in the past: scripts which bisect dependency version changes, Bayesian analysis of test}} \color{coolblack}{\small\textit{log messages to correlate with job success/failure, curated repositories of error conditions mined from software logs mapping to specific unresolved bug reports, databases of granular integration test success/failure within and across projects... the common thread is that tools \textbf{require maintenance and attention to their outputs}, i.e. dedicated time prioritized by contributors, a commodity in short supply for most open source projects (including OpenStack). Ultimately these tools, while fun to create, have fallen into disrepair and been abandoned due to \textbf{lack of enough people caring for them over the long term}.''}}
\end{formal}

$R_{6}$ noted that the absence of full-time paid developers focused on build processes may impede sustained tooling. 
\begin{formal}
    \color{coolblack}{\small\textit{There's full time [paid] people needed to consistently work on, fix and improve the [OpenStack] build process. I'm currently not aware of anyone doing this full time, [e.g.,] \textbf{most people spend a few hours per week on the CI, but there is no dedicated person working on it full time}, as far as I know.}}
\end{formal}

Five respondents also offered specific suggestions for tool support. $R_{5}$ proposed an error-tracking and monitoring tool such as Sentry,\footnote{\url{https://sentry.io/}} which is similar to the tools described by $R_{8}$. $R_{12}$ highlighted the value of enforcing consistent test environments across projects to deal with flakiness in general. $R_{7}$ recommended specialized tooling for event-related flakiness, while $R_{6}$ suggested verifying or invalidating build configurations across projects.

While participants such as $R_{15}$ agreed that rechecking practices could be improved with automation, $R_{8}$ raised concerns about blind rechecking, emphasizing that shared ownership benefits OpenStack overall:
\begin{formal}\color{coolblack}{\small\textit{``The CI/CD system OpenStack relies on (Zuul), is constantly evolving and improving, and is the most effective place we've found to collaborate on implementing or integrating what solutions and mitigations we do come up with. That's why it came into existence in the first place. I would encourage contributors to \textbf{pay attention to job results and learn to investigate failures they don't understand}. We've had far too many cases of \textbf{nondeterministic bugs introduced into projects because people just kept blindly rechecking} until they got lucky with successful results enough times in a row to get a [chance] to merge, at which point it becomes everyone's problem (including users when that nondeterministic behavior eventually ends up in their production deployments). Even if the failure seems to be unrelated to the project your change is touching, fixing bugs in OpenStack benefits OpenStack as a whole, which includes that sub-component project.''}}\end{formal}



\section{Threats to Validity}
\label{sec:tv}
Below, we describe the threats to the validity of our study.

\textbf{Construct Validity.}
To determine if flaky jobs occur across multiple projects (RQ1) and, at a more granular level, whether a flaky test exhibits different levels of flakiness (RQ2), we analyze the logs of 1,651 unique Zuul jobs. As such, it is possible that jobs and tests may be misclassified due to insufficient \fix{collection, e.g., executions of a flaky test are not captured in specific projects.} To combat this threat, we collect at least 200 runs for each flaky job\fix{, obtaining 87,711 valid test result files,} which reduces the likelihood of such types of misclassification. \fix{In RQ2, our test scope classification is based on community-defined labels embedded within the dot-delimited fully-qualified class name of tests~\cite{opendoc,opendoc-unit,opendoc-fullstack}, rather than strict definitions of test scopes. The choice is determined to reflect the practical observation in the OpenStack ecosystem.}

\textbf{Content Validity.} 
This study does not manually classify the entire data set \fix{or a fixed-size random sample} of flakiness inconsistencies, posing a risk of undiscovered cause categories. Instead, we strive for theoretical saturation \cite{eisenhardt1989building} to achieve analytic generalization, a broadly adopted approach in software engineering research \cite{hirao2019review,rigby2011understanding,xiao2021characterizing}. To combat this threat, we conduct \fix{eight iterations and achieve saturation after coding 110 inconsistently flaky tests and reviewing 671 flaky instances to triangulate CI logs, error messages, and bug reports}.

\textbf{Internal Validity.}
\fix{In RQ2, we investigate the relationship
between test scopes and flakiness levels. The observed distributions may be attributed to the disproportionately high test or execution frequency of specific test scopes (e.g., unit tests) as well as disparities in CI job configurations across the OpenStack ecosystem. Given the difficulty of tracking the total volume of test executions across the entire OpenStack ecosystem, we report the average number of executions for each test scope. These values are derived from occurrences in our data set and may not accurately represent executions across the OpenStack ecosystem. Furthermore, we also mitigate this potential bias by reporting relative numbers rather than absolute numbers for Observation 5.} 

Since the causes of inconsistent flakiness rely on manually coded data in RQ3, potential miscoding may arise due to the subjective nature of schema interpretation and experience in OpenStack, as well as in searching for test symptoms on their bug report system. To mitigate this, we conduct a developer questionnaire and our responses of the respondents support these causes (Table~\ref{tab:sur}). In addition to those we identified in RQ3, only one other potential cause was recognized: text fixtures originating from diverse vendors. \fix{While we validate these causes in the developer questionnaire, we did not cross-validate them across the broader OpenStack ecosystem nor perform a granular, case-by-case validation with OpenStack developers. Consequently, there is a risk that specific instances may have been misattributed to incorrect bug reports. Moreover, without performing cross-validation, we can not distinguish which causes were the true behavioral inconsistency that may be crucial for repairing inconsistent flakiness. Future work could partner with OpenStack maintainers to identify and validate true behavioral
inconsistency of inconsistent flakiness.}

\textbf{External Validity.} Our study focuses exclusively on the OpenStack ecosystem, which may limit the applicability of our results to other ecosystems. We opt to emphasize internal validity by analyzing 94\% of OpenStack projects that employed Zuul CI, providing substantial coverage within this domain. 

\fix{To mitigate it, we examine the analytic generalizability~\cite{yin2009case} of the two identified phenomena (cross-project flakiness and inconsistent flakiness) by extending our analysis to another ecosystem, Zuul, a widely adopted \minor{CI/CD system. Although Zuul was originally developed for OpenStack projects, it has since become independent of OpenStack, with open-source and proprietary users outside of the OpenStack ecosystem.\footnote{https://zuul-ci.org/users.html} Thus, Zuul forms its own, albeit smaller, ecosystem, which serves as a valuable case to validate the phenomena we identified in this paper. Moreover, Zuul CI is also integrated into the code review process of the Zuul ecosystem, allowing us to reuse the same approach outlined in Sections~\ref{sec:cs}--\ref{sec:rq2}. In total, we collect 757 closed Zuul reviews across six projects from January 2025 to November 2025.}
For cross-project flakiness, we identify one job (\path{zuul-quick-start}) out of 53 Zuul jobs exhibiting non-deterministic behavior across multiple projects. For inconsistent flakiness, we found three flaky tests out of 51. Moreover, we discover that 46 of them are unit tests, which is consistent with Observation 3.
These results collectively confirm the applicability of our results in Zuul. 
In future work, we plan to extend the investigation to other \minor{larger} ecosystems to further enhance generalizability.}

\section{Related Work}
In this section, we position our work with respect to the literature on flaky test identification, Continuous Integration (CI) in modern code review, and flaky tests in CI.

\textbf{Flaky Test Identification.}
Re-running tests is often employed to pinpoint tests that non-deterministically pass and fail on the same code. Tools such as DeFlaker~\cite{bell2018deflaker} leverage differential code coverage to detect failing tests that did not execute newly changed code (likely flaky failures). Similarly, iDFlakies~\cite{lam2019idflakies} reruns tests in different orders to discover both order-dependent and non-order-dependent flakiness. These automated rerun detectors uncover flaky tests that may not be caught in a single attempt. 

Due to the time-consuming nature of reruns, researchers explored static analysis and machine learning for predicting flaky tests. The FlakeFlagger tool~\cite{alshammari2021flakeflagger} uses test smells and metrics to predict flakiness without test execution, attaining an F1 score of 86\%. Flakify~\cite{fatima2022flakify} uses a pre-trained model and source code features to achieve impressive cross-validation performance, being especially adept at identifying flaky tests in unseen projects. While these models are promising, all flaky tests cannot be identified using code structure alone.

Research on Pull Requests (PRs)~\cite{lam2020study}, commits~\cite{luo2014empirical}, and bug fixes~\cite{7332456} has been conducted to explore flaky tests in practice. For example, Lam et al.~\cite{lam2020study} examined PRs that fixed flaky tests in six Microsoft projects, observing recurrent issues like asynchronous waits or race conditions, and incomplete fixes. This indicates that identifying flakiness is challenging; developers may believe that a test is stable after changes, but the test remains flaky. It warns that relying solely on developers' reports of flaky test fixes can be misleading. Overall, identifying flaky tests often requires combining multiple signals (rerun results, static code patterns, historical failure data) to differentiate genuine failures from false alarms.

\textbf{CI in Modern Code Review.}
Rahman and Roy~\cite{rahman2017impact} analyzed the impact of automated builds on
code reviews, observing that passing builds had a large influence on code review participation. 
To understand how developers use the outcome of CI
builds during code review, Zampetti et al.~\cite{zampetti2019study} empirically investigated the interplay between PR discussion and the use of CI.
Their study indicated that successful CI builds positively affect code review results compared to failed ones, while questionnaire findings highlighted that CI adds complexity to code review configuration and maintenance. Bernardo et al.~\cite{bernardo2018studying} found that projects deliver merged PRs more
quickly after adopting CI, and increasing the number of PRs submitted for projects slows their delivery time. Other researchers also found that: (i) adopting CI could reduce commenting noise as a silent helper~\cite{cassee2020silent}; (ii) difficulty in resolving CI failures is one of the reasons contributors abandon their PRs~\cite{khatoonabadi2023wasted}; (iii) longer CI build time can prolong the review duration, but passed CI could be reviewed in a shorter time.

\textbf{Flaky Test in CI.}
In CI environments, flaky tests pose challenges, and past studies have explored CI data to detect such tests. One method observes restarted builds; if developers rerun a failing build without modifying the code and it passes, a flaky CI step is to blame. An empirical study~\cite{durieux2020empirical} of Travis CI found developers manually restarted 1.72\% of all builds, totaling over 56,000 restarts, with nearly half succeeding upon retries. Maipradit et al.~\cite{maipradit2023repeated} shifted focus to OpenStack, where the ``recheck'' command allows reruns of CI jobs in code review, showing that 55\% of OpenStack reviews involved repeated builds, with 42\% altering test results, exemplifying flakiness. These rechecks consumed the equivalent of 187.4 CPU years and delayed code review decisions by a total of 16.8 years, excluding justified repeats. Similar issues exist in proprietary CI, such as at Google~\cite{googleblog1}, where 1.5\% of tests are flaky and 16\% show flakiness. At Google, 84\% of post-submit test failures (pass to fail), were due to flakiness instead of code changes, meaning five out of six such failures were false alarms. This wastes failure diagnosis effort, desensitizing developers who may overlook genuine issues. Olewicki et al.~\cite{olewicki2022towards} reported that 31\% of the commits had at least one manually rerun build job, and 15\% of the build jobs were rerun at least once at one of the world's leading AAA game producers. \fix{Leinen et al.~\cite{leinen2024cost} discovered that at least 2.5\% of the productive
developer time is spent on flaky tests in industrial CI; however cost for rerunning tests is negligible to the developers' efforts.} Studies in open source and industry CI settings have demonstrated that flaky tests incur high costs through wasted resources, delayed actions, and time spent diagnosing unreliable test signals.

Unlike prior work, our study focuses on flakiness that manifests code review. We complement the prior work that has focused on open-source CI services (e.g., Travis CI~\cite{durieux2020empirical}) and proprietary CI systems (e.g., Google's internal pipeline~\cite{googleblog1}) by studying an infrastructure-based open-source ecosystem (OpenStack). We also adopt a more fine-grained perspective.
Prior work on restarted builds in Travis CI examined how flaky builds prolong code review time, whereas we quantify the additional build time incurred by these flaky builds. Similarly, studies of repeated builds in OpenStack~\cite{maipradit2023repeated} have often been coarse-grained, treating each rerun as a single unit, In contrast, our analysis is fine-grained, drilling down to individual failing tests within repeated failing builds. Finally, unlike earlier research that primarily investigated root causes of flaky tests, we broaden the scope by examining cross-project flakiness (flaky tests that affect multiple projects) and inconsistent flakiness (where a test behaves non-deterministically in some projects but remains stable in others).

\section{Conclusion and Lessons Learned}
Automated regression testing has become an essential component of modern code review and CI workflows. In this context, test flakiness not only disrupts individual projects, but also introduces complex challenges at the ecosystem level. In this paper, we identify and analyze cross-project flakiness and inconsistent flakiness. Below, we distill the lessons for developers, team leads, and researchers of software ecosystems.

\textbf{Suggestions for Developers.}
\begin{itemize}
    \item Automate dependency validation: Our results show that inconsistent flakiness may stem from dependency issues with the patch (Observation 6). Therefore, automating dependency compatibility checks is crucial to verify that all libraries and tools across projects are version-compatible before integration ($R_{12}$). This proactive management of dependencies will mitigate instances of flakiness, particularly in multi-project setups. 
\item Adopt a preventative rather than reactive approach:
More than 50\% of OpenStack projects experience cross-project flakiness when rechecking failing builds (Observation 1). Non-cross-project flakiness leads to time waste, while cross-project flakiness generates even more time waste (Observation 8). This could be attributed to the ``recheck and wait'' paradigm prevalent in OpenStack. Therefore, we suggest that developers shift from a ``recheck and wait'' paradigm to early intervention in flaky tests and repeated failing builds. We encourage developers to examine even seemingly unrelated failures can prevent non-deterministic bugs reaching production, as $R_8$ explained ``people just kept blindly reaching until they got lucky [...] to merge, [then] it becomes everyone's problem. [fixing these bugs] benefits OpenStack as a whole.''
\item Automate flakiness classification:
We classified the root cause of inconsistent flakiness into three main categories (Observation 6), each has a unique value. We recommend extending CI tooling to classify flaky tests by their root causes ($R_6$ and $R_7$) \fix{based on the historical logs, failure symptoms, and environments, which are used to detect flaky tests in CI~\cite{lampel2021life, alshammari2024230, haben2024importance}}. This will aid in prioritizing repair efforts effectively, e.g., event-related flakiness is likely due to external events, thus making it a low-priority task for manual log inspection and repair. For example, Brus et al.~\cite{brus2025characterizing} propose a recheck prediction model that will automatically recheck a failing build when the predicted likelihood of passing is high. \fix{Furthermore, the root causes of cross-project and inconsistent flakiness may differ: the former can arise from intrinsic test logic~\cite{eck2019understanding}, whereas the latter may be driven by ecosystem-level factors. A systematic classification that distinguishes between these two types of flakiness, as well as between the contexts of single-project and cross-project, could be beneficial.}
\end{itemize}

\textbf{Suggestions for Team Leads.}
\begin{itemize}
\item Redesign reusable unit tests:
An unexpected finding is the occurrence of flakiness in unit tests, which are traditionally designed to validate isolated components (Observations 3 and 5). This suggests a misalignment
between the test design and its scope, particularly when unit tests
are reused across multiple projects. Therefore, we suggest weighing the benefits of isolation testing, such as dependency injection, mocking, or stubbing, which reduce flaky dependencies on shared resources and configurations, against the potential trade-off of creating tests that are less realistic of the production environment.
\item Centralize tracking of test flakiness: Cross-project flakiness often occurs in the \fix{ecosystem-level} CI workflow (Observations 1 and 2), and
the results of our questionnaire ($R_5$, $R_8$, and $R_{15}$) indicate the benefit of creating or improving a shared repository for tracking incidences of flakiness. This resource would aid in failure diagnosis and coordination of debugging effort across projects within the ecosystem. While OpenStack provides OpenSearch (i.e., a community-maintained cluster that monitors failing tests and jobs), it requires continual upkeep and active community engagement to remain viable ($R_8$).
\item Routinely revisit CI configuration and environment:
Inconsistent build configuration is a primary cause of flakiness, showing a deterministic status in other projects (Observation 6). We suggest that team leads schedule routine reviews and updates in all projects to ensure that configurations remain aligned as new practices or requirements emerge ($R_{12}$). Encouraging containerization or standardized testing environments across projects could mitigate environment-specific flakiness and reduce debugging time.
\item \fix{Encourage multiple team collaboration and make invisible work visible: Reflecting the perspectives of $R_{5}$ and $R_{10}$, cross-project flakiness in the OpenStack ecosystem emerges as a socio-technical failure that necessitates coordinated effort across multiple teams. In particular, contributors (including newcomers) who experience unresponsive or delayed communication become discouraged from engaging in maintenance activities and instead adopt a ``recheck and wait'' paradigm. Consequently, we recommend that team leads actively foster inter-team collaboration to address cross-project flakiness and make these currently invisible maintenance efforts visible to the broader community, supported by appropriate incentives.}    
\end{itemize}

\textbf{Suggestions for Researchers.}
\begin{itemize}
    \item Examine long-standing flakiness-related bug reports: In the open coding process of RQ3, we found that developers usually use inline comments (``recheck bug \#XXXXXXX'') to trigger the recheck using a persistent flakiness-related bug report. We recommend that researchers examine these bug reports (which can be captured by terms such as ``unstable,'' ``flaky,'' and ``non-deterministic'' on Launchpad). This 2019 bug report,\footnote{\url{https://bugs.launchpad.net/mistral/+bug/1812171}} remains unassigned, yet it is categorized by the bug supervisor as ``Triaged'' (indicating all necessary fix information is present) and is of medium importance.
    \item Analyze the long-term impact of cross-project flakiness:
Cross-project flakiness generated more time waste compared to single-project flakiness (Observation 8); we suggest that researchers conduct longitudinal studies to \fix{explore its effects on long-term maintenance and software quality, such as the effectiveness of test suites over time} in CI and propose strategies to mitigate these impacts.
\end{itemize}


\section*{ACKNOWLEDGMENTS}
We thank all participants who took the time to complete our questionnaire, providing valuable insights for our research. We gratefully acknowledge the financial support of: (1) JSPS for the KAKENHI grants (25K22845, 23KJ1589, 26H02500, and 26K21198); (2) Japan Science and Technology Agency (JST) as part of Adopting Sustainable Partnerships for Innovative Research Ecosystem (ASPIRE), Grant Number JPMJAP2415; (3) JST PRESTO, Grant Number JPMJPR22P6; (4) the Kayamori Foundation of Informational Science Advancement for supporting Tao Xiao; (5) the Inamori Research Institute for Science for supporting Yasutaka Kamei via the InaRIS Fellowship; and (6) NSERC for supporting Shane McIntosh via Alliance International Collaboration Grant ALLRP 580835 - 22.

\bibliographystyle{IEEEtranS}
\bibliography{main}

\end{document}